\documentclass[aps,prd,amsmath,amsfonts,  a4paper,11pt,reprint,twocolumn,square,numbers,showpacs,
floatfix,sort&compress]{revtex4-1}

\usepackage{hyperref}
\usepackage{amssymb}
\usepackage{bm}
\usepackage{natbib}
\usepackage{graphicx}
\expandafter\let\csname equation*\endcsname\relax
\expandafter\let\csname endequation*\endcsname\relax
\usepackage{amsmath}
\usepackage[labelfont={bf,scriptsize},textfont={scriptsize},justification=RaggedRight,format=hang]{caption,subfig}
\usepackage{color}

\RequirePackage{ifpdf}
\ifpdf
\else 
\fi

\usepackage{titlesec}
\titlespacing*{\section}{0pt}{0.6cm}{0.5cm}

\usepackage{soul}
\definecolor{dm}{cmyk}{.0, .6, .90, 0}

\sethlcolor{dm}

\newcommand{\Mpl}{M_{\mathrm{Pl}}}

\newcommand\be{\begin{equation}}
\newcommand\ee{\end{equation}}
\newcommand{\bea}{\begin{eqnarray}}
\newcommand{\eea}{\end{eqnarray}}

\begin{document}

	\title{The Swampland, Quintessence and the Vacuum Energy}
	\author{M.C.~David Marsh}%
	\affiliation{Department of Applied Mathematics and Theoretical Physics,  University of Cambridge,
Cambridge, CB3 0WA, United Kingdom \\ \emph{Email:}~m.c.d.marsh@damtp.cam.ac.uk}
		
	\begin{abstract}
	\noindent
It has recently been conjectured that string theory does not admit de Sitter vacua, and that quintessence explains
the current epoch of accelerated cosmic expansion. A proposed, key prediction of this scenario is time-varying couplings in the dark sector,  induced by the evolving quintessence field. We note that cosmological models with varying couplings suffer from severe problems with quantum corrections, beyond those shared by all quintessence models. The vacuum energy depends on all masses and couplings of the theory, and even small variations of parameters can lead to overwhelmingly large corrections to the effective potential. We find that quintessence models with varying parameters can be realised in consistent quantum theories by either: 1) enforcing exceptional levels of fine-tuning; 2) realising some unknown mechanism that  cancels all undesirable contributions to the effective potential with unprecedented accuracy; or 3) ensuring that the  quintessence field  couples exclusively to very light  states, and does not backreact on heavy fields.
	\end{abstract}
	
	\maketitle

\section{Introduction}
An important question in fundamental physics is what distinguishes
general effective field theories from those that can be consistently realised in  quantum gravity. Inspired by examples of compactifications from string theory, the authors of \cite{Obied:2018sgi} conjectured that 
quantum gravity 
 severely restricts
the effective scalar potential, $V$, of the low-energy theory:
\be
|\nabla V| \geq c\, V \, ,
\label{eq:cond1}
\ee
for a positive constant $c\sim {\cal O}(1)$ and in units where $\Mpl = 1/\sqrt{8\pi G} =1$. 

If true, equation \eqref{eq:cond1} has far-reaching implications \cite{Obied:2018sgi, Agrawal:2018own, Andriot:2018wzk, Denef:2018etk,  Conlon:2018eyr, Cicoli:2018kdo, Akrami:2018ylq}. Most notably, equation \eqref{eq:cond1}  forbids local de Sitter  critical points (see also \cite{Danielsson:2018ztv}) and forces the current period of accelerated expansion to be realised through particular models of quintessence \footnote{
The implications of equation \eqref{eq:cond1} for inflation in the early universe were discussed in \cite{Agrawal:2018own, Achucarro:2018vey, Garg:2018reu, Kehagias:2018uem, Dias:2018ngv, Kinney:2018nny, Ben-Dayan:2018mhe, Matsui:2018bsy}.
}. Reference \cite{Agrawal:2018own} argued that such models can be naturally realised in string theory where slowly rolling moduli fields can support the accelerated expansion. 

Some well-known restrictions on quintessence  were discussed in \cite{Agrawal:2018own, Denef:2018etk,  Chiang:2018jdg,
Cicoli:2018kdo, Akrami:2018ylq}. Very light scalar fields coupled to the Standard Model can mediate long-range forces, which are severely constrained by precision tests of the equivalence principle. Moreover, scalar fields that modify the masses and couplings of the Standard Model are constrained by astronomical observations. Finally, models of quintessence require not only that the value of the scalar potential is very small, but so must its gradient.

In reference \cite{Agrawal:2018own}, the absence of observed variations in the Standard Model parameters were interpreted  as evidence for comparatively stronger  couplings between the quintessence scalar and some fields in the dark sector. 
This is not a direct consequence of equation \eqref{eq:cond1}, but is arguably natural as such a scenario 
can be realised in string theory through branes, e.g.~of type IIB or F-theory.
%
%
 For example, the quintessence field may control the volume of the cycle where dark matter originates, so that its evolution leads to variations in dark matter couplings. 
 In the cosmology literature, models realising dark energy/dark matter interactions are usually referred to as  `interacting dark energy' \cite{Wang:2016lxa}.


The purpose of this note is to recall that a cosmic scalar field, $\phi$,  that cause variations in couplings and masses suffer from {\it severe problems} when considered in quantum field theory \cite{Banks:2001qc, Marsh:2016ynw, Donoghue:2001cs, DAmico:2016jbm}.  The basic argument (reviewed in detail below) is that small variations in couplings cause large variations in the vacuum energy. For example, a variation in a fine-structure  constant $\alpha(\phi) = \bar \alpha + \delta \alpha$ to which matter with large mass $M$ is coupled leads to a variation of the vacuum energy that is schematically of the form,
\be
\delta \rho_{\rm vac} \sim \delta \alpha(\phi)\, M^4 \, .
\ee
This is a contribution to the low-energy effective potential of $\phi$ that can overwhelm any naive quintessence potential. 
This makes it very challenging to promote cosmological models of varying `constants' into consistent quantum theories.




In this note, we apply these arguments to the recently proposed quintessence models of \cite{Agrawal:2018own}, and find that they can only be realised 
under certain restrictive conditions.


\section{The vacuum energy and varying parameters}
The one-loop Coleman-Weinberg potential for a general field theory in four-dimensional  flat space is given by \cite{Coleman:1973jx, Weinberg:1973ua, Iliopoulos:1974ur},
\bea
\delta V &= \frac{1}{(8\pi)^2} \left[
\Lambda^4\, {\rm STr}(M^0) \ln\left( \frac{\Lambda^2}{\mu^2}\right) + 2 \Lambda^2\,  {\rm STr}(M^2) 
\right.\nonumber \\
&\left.
+ {\rm STr}\left( M^4\right) \ln \left(\frac{M^2}{\Lambda^2}\right)
+ \ldots
\right] \, ,
\label{eq:CW}
\eea
where $\mu$ is scale parameter and $\Lambda$ the cut-off scale. The supertrace is given by  ${\rm STr}(M^n) = \sum_i (-1)^{2j_i} (2j_i +1) m_i^n$ where $j_i$ is the spin of the different particles with mass eigenvalues $m_i$. 
The first term is always field-independent, vanishes for spontaneously broken supersymmetric theories,  and is only relevant for the original cosmological constant problem. In spontaneously broken supergravities, the supertrace is generically non-vanishing for $n>0$, but in some special `no-scale' supergravities, the $n=2$ term can vanish even after supersymmetry breaking \cite{Ferrara:1994kg}. In this note, we will conservatively consider only the third term, which is only logarithmically sensitive to the model-dependent cutoff.





Including also  higher loop-order corrections, we may write these contributions as,
\be
\delta V = \frac{1}{(8\pi)^2} \sum_i c_i\, m_i^4  + \ldots\, , 
\label{eq:vac}
\ee
where the coefficients $c_i(\alpha)$ depend on the coupling constants of the theory, and absorb any logarithmic factors. Accounting for loop-factors, $\partial_{\alpha}^p c_i \sim  {\cal O}((4\pi)^{-p})$. The effective quintessence potential below the scale $\Lambda$ is then given by the sum of the bare contribution $V_0(\phi)$ and the loop corrections: $V = V_0 + \delta V$.


It is now easy to understand the particular problems associated with quintessence models with varying parameters. A change in the Standard Model fine-structure constant of the order of the current observational limit, $\delta  \alpha/ \alpha \sim10^{-6}$, leads to a change in the vacuum energy  of the order of (cf.~\cite{Banks:2001qc}),
\be
\delta V \sim (170\, {\rm MeV})^4 = 3 \times 10^{43}\, \rho_0 \, ,
\ee 
where we have taken ${\rm Max}(m_i)=m_{\rm top} = 173$ GeV and the vacuum energy is $\rho_0 = (2.3\times10^{-3}\, {\rm eV})^4$.
Since $\alpha$ is field dependent, this is a highly disruptive contribution to the effective potential of $\phi$ \footnote{We note that if $\phi$ only has couplings that respect the shift symmetry $\phi \to \phi+{\rm constant}$, the variation of $\phi$ will cause no change in coupling constants of the theory. }.

\section{The flatness of the quintessence potential}
Even if the quintessence is completely decoupled from the Standard Model, small changes in the parameters of the dark sector can lead to overwhelmingly large contributions to the quintessence effective potential. 
An important question is then: {\it 
given the vacuum energy contribution of equation \eqref{eq:vac}, how can the quintessence potential be sufficiently flat? }

Excessive fine-tuning is one option. In order for the potential to be sufficiently flat, not only the value of the potential and its gradient need to be tuned, but also higher-orders  in the Taylor expansion around the present-day value. 
Suppose that an evolution of the quintessence field by $\delta \phi$ causes a variation, $\delta \alpha _{\rm tot}$, in a dark sector coupling constant under which matter of mass $m_i$ is charged. 
Imposing that the value of the potential over this range does not exceed $\rho_0$ then  requires cancellations of the $k$:th order in a Taylor expansion to at least one part in \cite{Marsh:2016ynw},
\be
\left( \frac{m_i^4}{(8\pi)^2 \rho_0}\right) \left(\frac{ \delta \alpha_{\rm tot}}{4\pi}\right)^k \, , 
\ee  
for $k\leq k_{\rm max} = {\rm floor}\left[\ln(\frac{m_i^4}{(8\pi)^2 \rho_0}))/\ln \left(\frac{ 4\pi}{\delta \alpha_{\rm tot}}\right) \right]$.
For example, if the dark matter has mass $m= 100\, {\rm GeV}$ and is charged under the gauge group with a coupling constant that changes by $\delta \alpha_{\rm total}  = 10^{-2}$,  the quintessence potential need to be tuned up to  order $k_{\rm max} = 16$. 
The total fine-tuning (on top of that required by the original cosmological constant problem) is 
the product of the required fine-tuning of the individual operators and is given by \cite{Marsh:2016ynw}:
\be
{\mathfrak f} = \left( \frac{m^4}{(8\pi)^2 \rho_0}\right)^{\frac{1}{2}(k_{\rm max} -1)}  = 1 \times 10^{388} \, .
\ee
In the absence of this tuning, $\phi$ could get stuck in a de Sitter minimum or rapidly evolve towards a crunch.

A second option is that a new symmetry or mechanism cancels all undesirable contributions to the potential with exceptional accuracy. Unbroken global supersymmetry sets ${\rm STr}(M^n)=0$ \cite{Zumino:1974bg}, but this is no longer true in supergravity \cite{Grisaru:1982sr, Ferrara:2016ntj}.
A new cancellation mechanism may be intrinsically string theoretic and 
appear
unexpected from a supergravity viewpoint. While this is an intriguing possibility,  no such mechanism has yet been identified in the low-energy theories arising from string compactifications. 
The discovery of such a mechanism through careful string theory calculations would 
strengthen the case for 
 the cosmology proposed in \cite{Agrawal:2018own}, and possibly  for 
the correctness of equation
\eqref{eq:cond1}.

A third option is that $\phi$ couples exclusively to very light states, so that the  equation \eqref{eq:vac} gives a  negligible contribution to the  effective potential of $\phi$ (cf.~\cite{DAmico:2016jbm} for an example). This may be realised rather naturally if $m_i^4(\phi) \ll V_0(\phi)$  as $\phi$ descends the quintessence potential. In the present era, such fields should be no more massive than $m_i \lesssim 4\times 10^{-2}\, {\rm eV}$ to allow for ${\cal O}(1)$ changes in parameters without spoiling the quintessence potential. 

This solution is comparatively appealing, but has two important caveats. First,  to convincingly realise such a mechanism one must demonstrate that the contribution from the second term of the Coleman-Weinberg potential \eqref{eq:CW}  is negligible. If the cutoff $\Lambda$ is close to the Planck scale, this may require the stricter limit $m_i \lesssim {\cal O}(H_0)$, in which case $\phi$ can only couple to other quintessence fields. 

Second, the parametrically large hierarchy between particle physics mass scales and the vacuum energy requires that  $\phi$  interacts extremely weakly with other moduli. 
For example, if the evolution of $\phi$
 changes  the total volume of the compactification or the string coupling constant, the spectrum of massive states will change, and the vacuum energy problem is re-introduced.  
 
 To illustrate how sharp this decoupling must be, suppose that a Standard Model   gauge coupling, $g$, is controlled by a volume modulus, ${\cal V}_{\rm SM} = 1/g^2$.
 For concreteness, we take $\alpha =1/25$, as is appropriate for Grand Unified Theories. 
 As $\phi$ evolves, ${\cal V}_{\rm SM}$ must stay fixed to a high accuracy, or the Standard Model vacuum energy  corrections dominate over the quintessence potential. This requires,
\be
\frac{\delta {\cal V}_{\rm SM}}{{\cal V}_{\rm SM}} < 6\times 10^{-51} \, ,
\ee
where we have again set $m=m_{\rm top}$ and require $\delta V_{\rm SM} < \rho_0$.
Such a rigidity of the Standard Model cycle can be challenging to realise when all fields are (at least gravitationally) coupled, and $\phi$ evolves substantially. 

In closing, we recall that 
the conjecture \eqref{eq:cond1} is violated 
if the potential for $\phi$ is additively combined with 
the  Standard Model Higgs potential, $V_H = \lambda_H \left( |H|^2 - v^2\right)^2$, and evaluated at $H=0$ \cite{Denef:2018etk}.
After first identifying this issue, reference  \cite{Denef:2018etk} considered a simple modification of the coupling between the Higgs field and $\phi$ that avoids this problem:
\be
V_0 = e^{-\lambda \phi} \left(V_H(H) + \Lambda \right) \, .
\label{eq:mod}
\ee
We note that equation \eqref{eq:mod} leads to substantial variations in the Higgs sector parameters, and consequently to large quantum corrections to the potential.
These models must  then realise  either of the first two options identified in this paper to explain the 
 present-day accelerated expansion through quintessence.

\section{Conclusion}
The drastically simple condition  \eqref{eq:cond1} has been proposed to delineate the `swampland' of theories that cannot be embedded into any consistent theory of quantum gravity.  The current status of this conjecture is highly uncertain and controversial \cite{Andriot:2018ept, Andriot:2018wzk, Dvali:2018fqu, Denef:2018etk, Roupec:2018mbn, Conlon:2018eyr,  Dasgupta:2018rtp, Kachru:2018aqn, Cicoli:2018kdo, Akrami:2018ylq}, in particular as detailed calculations demonstrating the failures of apparent counter examples  
are still lacking. 

Equation \eqref{eq:cond1} excludes de Sitter vacua, but is compatible with  certain models of  quintessence. A key prediction of reference \cite{Agrawal:2018own} is that such models cause cosmological variations in the couplings of dark matter and other dark sector fields. In this note, we have considered the theoretical implications of this proposed cosmology, and we have shown that they suffer from severe quantum instability problems. Variations in the couplings of massive states lead to large contributions to the vacuum energy that must be cancelled to an incredible accuracy. 
This instability problem is distinct from the 
cosmological constant problem as well as the
regular fine-tuning problem of quintessence models. 

We have shown that if the quintessence models of \cite{Agrawal:2018own} are realised in nature, one out of three conditions must hold: 1) the theory is incredibly fine-tuned; 2) there is a new, fantastic mechanism that surpasses even supersymmetry in taming dangerous quantum corrections; or 3) the quintessence field couples only to light states.

These conditions severely restrict the realisations of these models in any quantum theory, including string theory.


\section*{Acknowledgements}
I am grateful to Joseph Conlon, Arthur Hebecker and Timm Wrase  for comments on a draft of this paper, and I acknowledge support from a Stephen Hawking Advanced Fellowship at the Centre for Theoretical Cosmology at the University of Cambridge.

	\raggedright
\bibliography{refs}

\end{document}